\documentclass[12pt, prd]{revtex4}
\usepackage{amsmath}
\usepackage{amssymb}
\usepackage[utf8]{inputenc}
\usepackage{graphicx,epsfig}
\usepackage{xcolor}

\begin{document}

\title{Einstein and Eddington and the eclipse in Principe: \\ Celebration
and science 100 years after}
\vspace{1cm}
\author{\vskip 0.4cm Jos\'{e} P. S. Lemos$^{1}
\footnote{joselemos@tecnico.ulisboa.pt}$,
Carlos A. R. Herdeiro$^{2}\footnote{herdeiro@ua.pt}$,
and Vitor Cardoso$^{1}\footnote{vitor.cardoso@tecnico.ulisboa.pt}$}
\affiliation{\vskip 0.3cm $^{1}$Centro de Astrof\'isica e Gravita\c{c}\~ao
- CENTRA, Departamento de F\'isica, Instituto Superior T\'{e}cnico -
IST, Universidade de Lisboa - UL, Avenida Rovisco Pais 1, 1049-001,
Lisboa, Portugal\email{joselemos@tecnico.ulisboa.pt,
vitor.cardoso@tecnico.ulisboa.pt}
\vskip 0.1cm $^{2}$Departamento de Matemática da Universidade
de Aveiro and Centre for Research and Development in Mathematics and
Applications - CIDMA,\\ Campus de Santiago, 3810-183 Aveiro, Portugal}

\bigskip

\begin{abstract}
\vskip 1cm
On May 29, 1919, at Roça Sundy,  Principe island,
Eddington confirms Einstein's general relativity theory
for the first time by photographing stars behind the obscured Sun
during a total eclipse.  History was made. At Sobral, Eddington's
astronomer colleagues photograph the same eclipse and also conclude
that light from distant stars suffers a  deflection when
passing by the gravitational field of the Sun,  in
accordance with general relativity.  With the confirmation of general
relativity, a theory of gravitation at a fundamental level, physics
became, once and for all, relativist and its future was
outlined.  The first world war had finished a few months before and
the deep wounds  between nations were yet to heal.
Science wanted to be above  it all showing that people
could  be united by a common goal. This year, the 100
years of this  achievement was commemorated with the
scientific conference ``From Einstein and Eddington to LIGO: 100 years
of gravitational light deflection'' in Principe to celebrate this
landmark event. We report here on this
conference of celebration.

\end{abstract}

\maketitle


\newpage

\section{The eclipse and the light deflection of 1919}

On May 29, 1919, an eclipse of fundamental importance
would take place. Stars behind a Sun hidden by the Moon could be seen due
to the deflection of light rays that passed in the gravitational field
of the Sun.

Einstein arrived at the final form of the theory of general relativity
in November 25, 1915, when he presented it to the Prussian Academy of
Sciences. In this theory, gravitational force was exchanged for
spacetime curvature. Moreover, he deduced that for light
incoming from distant stars grazing the Sun's surface,
the deflection of the light trajectory would be 1.75 arcseconds.
Newton's theory of gravitation gave half of that value, 0.875
arcseconds, and in addition there was the possibility of having no
light deflection at all in the case light would not couple to
gravitation.  Einstein had also showed that following his theory,
Mercury's perihelion would advance in accord with the astronomical
observations, this constituting an a posteriori proof of the theory, and
that light would suffer a spectral redshift when it would climb a
gravitational field, an experiment difficult to make. Thus, the
observation of the light deflection in a solar eclipse would be the
first direct proof of general relativity.

Eddington, a renowned astrophysicist from Cambridge  with a deep
knowledge of the theory, saw in the 1919 eclipse a supreme opportunity
to test Einstein's theory of gravitation, and convinced the English
astrophysicists, that in turn determined that it was high time to
test that prediction. The eclipse would pass along a track of 12
thousand  km from west to east  approximately along  the equator line.
Two expeditions, carefully planned by the royal astronomer, Frank
Dyson, left England in the beginning of March 1919, stopped in Lisbon
and then in Funchal where they parted. Eddington
 went to the island of
Principe with his assistant
Cottingham, a handicraftsman of clocks and other instruments.
They lodged at Roça Sundy, in the plantation house that
belonged to Jer\'onimo Carneiro and had all the necessary
infrastructures.  The Royal Greenwich Observatory astronomers,
Crommelin and Davidson, went to Sobral, installing the telescopes and
ceolostats in the horse track of the city's Jockey Club since there was no
race in the foreseeable future.

The eclipse of the Sun lasted 302 seconds, i.e., five minutes
and two seconds. With instruments functioning at their limits,
with better or worse weather, the two expeditions were a success,
managed to capture photographs of stars, for which the corresponding
light rays passed near the Sun, in plates that could constitute
the first direct proof of the theory of general relativity.
With the eclipse finished, the astrophysicists returned to England
to examine the collected images
through instruments that measured the displacements of stars in
photographic plates. Five months
after, the results revealed that the observed stars
near the solar disk during the eclipse were slightly shifted
in relation to their normal position in the sky, in the
amount predicted by Einstein's theory, i.e., 1.75 arcseconds
for stars near the Sun's rim.

The results were then announced on November 6, 1919, in a
meeting in the Royal Society jointly with the Royal Astronomical
Society. The observations had confirmed the theory of general
relativity and there was jubilation everywhere. The world now knew
that the correct theory of gravitation was not Newton's theory,
but instead general relativity, and Einstein turned into a celebrity
around the planet instantaneously. 
Physics, from this moment onward, became totally relativist,
one now knew that particles along with  their interactions,
including gravitation, obeyed without doubt the laws
of relativity. This is one of the most acclaimed events
in the history of science.

It was the beginning of a long and beautiful success story.  Black
holes, gravitational waves, and cosmology are natural, new, and major
consequences of the theory. One by one these consequences
were unravelled, with
general relativity passing in a magnificent manner a great number of
tests, the most recent and impressive being the direct detection by
LIGO (Laser Interferometer Gravitational-wave Observatory) of the
first gravitational wave in 2015. This wave, in turn, was generated by
the collision at cosmological distances of two black holes of  around 30 solar
masses each.
The theory has transformed in a fundamental way our understanding of
physics and astrophysics. It is also at the root of indispensable modern
technologies, as the Global Positioning System, or GPS for short,
only works with
the proper
synchronization between the clocks in the satellites and the
clocks on the Earth taking into account relativistic corrections.

Besides confirming the theory of general relativity, the May 29, 1919,
event showed once more that people from different nations could unite
to a common aim. At the time, the first world war had finished not
long ago, and English and German scientists, represented by Eddington
and Einstein, respectively, gave hands looking to a better future.

The physics and astrophysics worlds united anew this year of 2019
to praise and
celebrate this event. Given the historical character of this date,
several celebrations were organized.  Namely, in Principe, 100 years
after, there was a conference ``From Einstein and Eddington to LIGO:
100 years of gravitational light deflection'', that had the hallmark
of the Center for Astrophysics and Gravitation (CENTRA), a research
unit of Instituto Superior Tecnico (IST).  The action took place from
May 26 to May 30, and the stage was at resort Bom Bom, 3 km away from
Roça Sundy, the locus of Eddington's observations.  There was also
further celebrations in Principe, Sobral, Lisbon, Rio de Janeiro, and
London.

\section{The scientific conference in Principe in 2019:
Celebration of the history and the science}

In 2015 there were celebrations all around the planet
for the one hundred years of general relativity, commemorating
the publication by Einstein in November 25, 1915,
of the final and definitive form of the theory.
CENTRA having as a specific area of research the fundaments
of general relativity, celebrated this date
with a conference at IST ``GR 100 years in Lisbon'',
see~\cite{confsitegr100}.

Being the light deflection in the gravitational
field of the Sun a first
experimental test to general relativity after the theory was
elaborated, its historical verification with success
in the May 29, 1919, eclipse by Eddington and
collaborators, had to be celebrated. Without doubt the 1919
eclipse is one of the most acclaimed events in the
history of science and
of great significance for physics in general.  CENTRA, a center of
astrophysics and gravitation with works in experimental tests of
general relativity and other theories of gravitation as well,
did want to celebrate this iconic date.

The authors of this article considered thereby opportune and coherent
to link this notable confirmation to the scientific activities of
CENTRA and of other Portuguese scientists working in this area. Thus,
in December 2015, during the celebrations of the 100 years of general
relativity, we started conversations for a scientific conference in
the end of May 2019, to celebrate in turn the 100 years of the
confirmation of general relativity through the light deflection by the
Sun's gravitational field in the eclipse of May 2019, 2019.

As the observations were done in Principe and Sobral,
it would be natural that Portuguese scientists
would organize the scientific celebration in Principe,
a Portuguese territory at the time. The main objective of
the conference would be to celebrate this historical
date to reflect the legacy left by Einstein and Eddington
related to the eclipse and to discuss the
impressive subsequent developments on astrophysics and gravitation.
It would be a conference to tread history, 
share the extraordinary scientific advances, and 
to look into the future. The speakers would be chosen
along these ideas. Taking into account that the scientific
organization was from CENTRA, whose members have been developing
notable research in these areas and
are leaders or belong to leading international groups,
several CENTRA members would be chosen as speakers, together
with specialists from prestigious universities and institutions.

\begin{figure}[h]
\vskip -1cm
{\includegraphics[scale=0.65]
{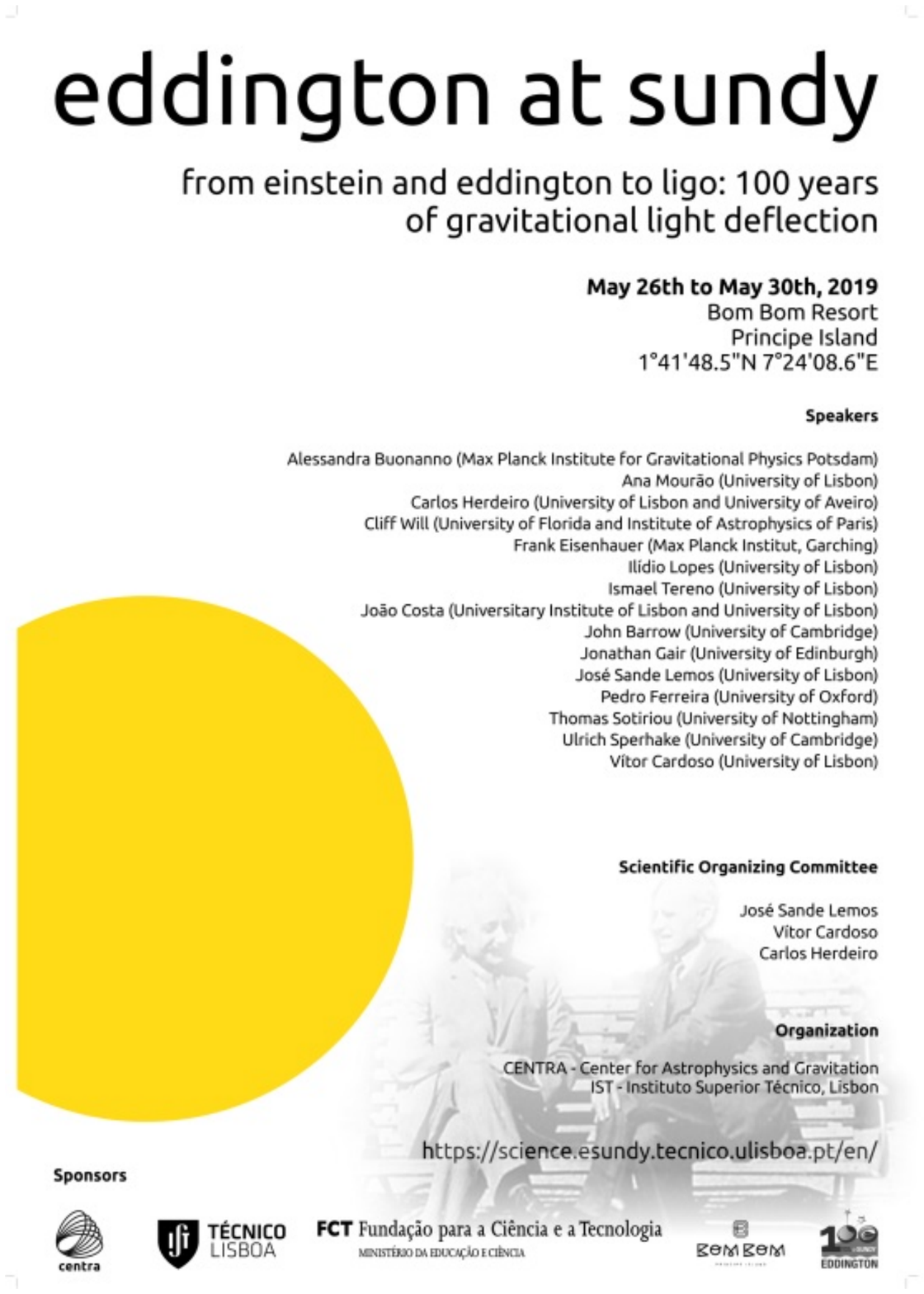}}
\vskip -3.5cm
\caption{\footnotesize The poster of the scientific conference ``From
Einstein and Eddington to LIGO: 100 years of gravitational light
deflection" in Principe.}
\label{poster}
\end{figure}

Having this in mind decisions were taken.
The chosen dates were May 26 to May 30, 2019, precisely
one hundred years after of the 1919 eclipse.
The  chosen venue in Principe for the scientific conference
was resort Bom Bom. It distances 3 km in a straight line and
9 km  by road to Roça Sundy, the place where
Eddington made the observations. 
Eddington writes in the eclipse article report that
when he and Cottingham arrived at Principe, after 
checking  the best place to install themselves,
they chose to mount the telescopes in Roça Sundy.
Curiously, the name Sundy is an English spell of
Sundi, that comes from Sumdim, that in the local language
means Senhor Dias, a  local land owner in the beginning of
the 19th century.
Within  the topics of
astrophysics and gravitation, it was established to
focus on themes related to the confirmation of light deflection
in a gravitational field and 
current themes at the frontier of general relativity,  such as
black holes, gravitational waves, and cosmology.
That is why the conference title
``From Einstein and Eddington
to LIGO: 100 years of gravitational light deflection" 
was chosen.  
In relation to the speakers, members of CENTRA together with
specialists and researchers in universities and institutes of prestige
were selected.  The speakers were fifteen, namely, Alessandra Buonanno
from the Max Planck Institute in Potsdam, Ana Mourão from the
University of Lisbon, Carlos Herdeiro from the University of Lisbon
and University of Aveiro, Clifford Will from University of Florida and
University of Paris, Frank Eisenhauer from the Max Planck Institute in
Garching, Ilídio Lopes from the University of Lisbon, Ismael Tereno
from the University of Lisbon, João Costa from the University
Institute of Lisbon, John Barrow from the University of Cambridge,
Jonathan Gair from the University of Edinburgh and from the Max Planck
Institute in Potsdam, José Sande Lemos from the University of Lisbon,
Pedro Ferreira from the University of Oxford, Thomas Sotiriou from the
University of Nottingham, Ulrich Sperhake from the University of
Cambridge, and Vítor Cardoso from the University of Lisbon.  The
conference  webpage was put online~\cite{confsite2} and the conference
poster was made public, see Fig.~\ref{poster} and~\cite{poster} .

The speakers and participants  arrived on May 26 in Principe
after one day stop over in São Tomé and it was with enormous
jubilation that we all have celebrated during the conference
in Principe the 100 years of this historical eclipse.
Resort Bom Bom situated by the sea shore 
in a marvellous place of the island with paradisiac beaches,
has a seminar room surrounded by equatorial vegetation, inspiring for
this conference of celebration.
May 27 and 28 were dedicated to talks,   in May 29
the stage was at Roça Sundy.

\begin{figure}[h]
{\includegraphics[scale=0.20]{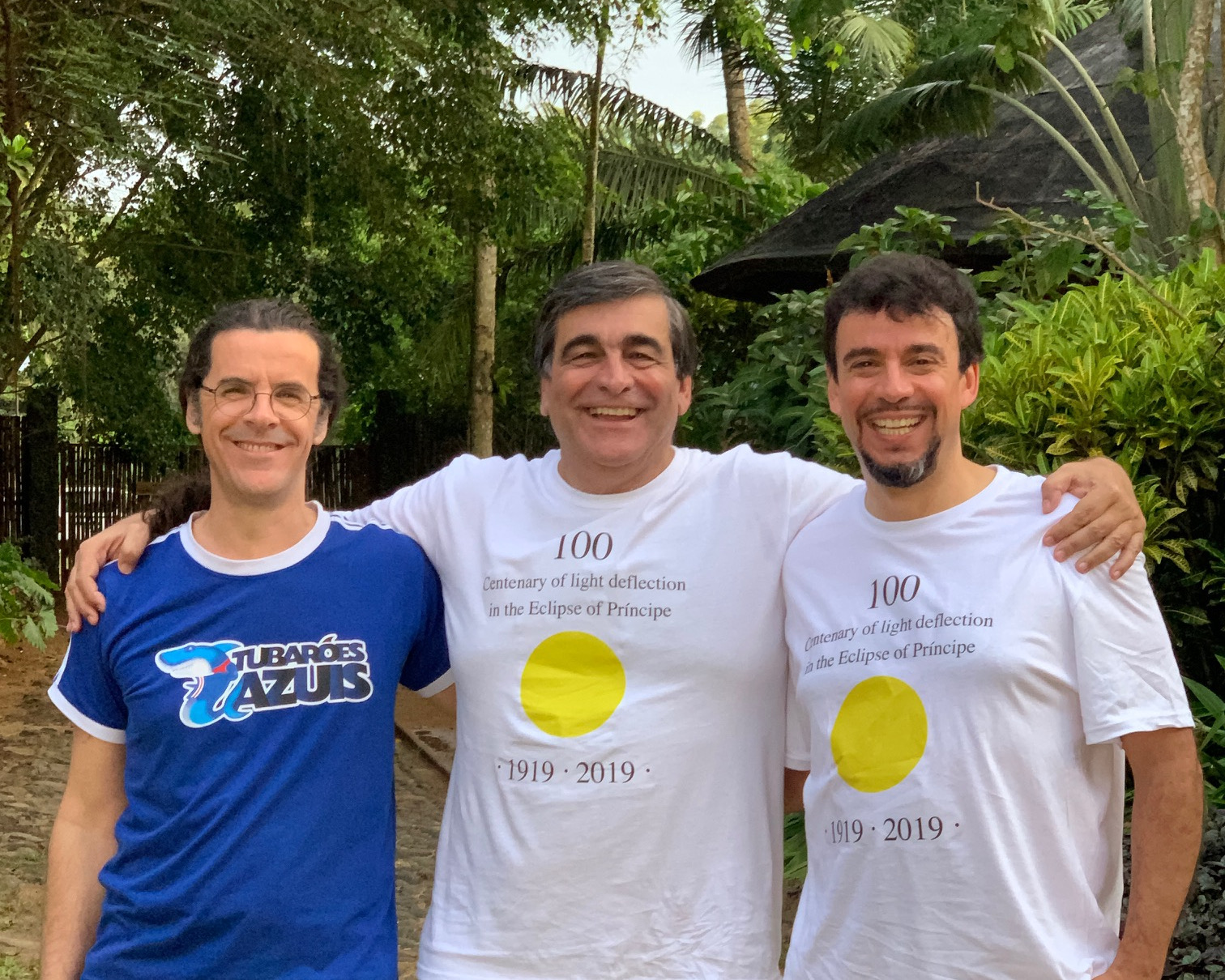}}
\caption{\footnotesize The organizers of the scientific
conference 
``From Einstein and Eddington to LIGO: 100 years of gravitational
light deflection'' in the middle of the luxuriant
vegetation in resort Bom Bom,
Principe. From left to right: Vitor Cardoso, José Sande
Lemos, Carlos Herdeiro. Photograph taken by Ilídio Lopes
in the morning of May 27, 2019.}
\label{organizers}
\end{figure}

\begin{figure}[h]
{\includegraphics[scale=0.18]{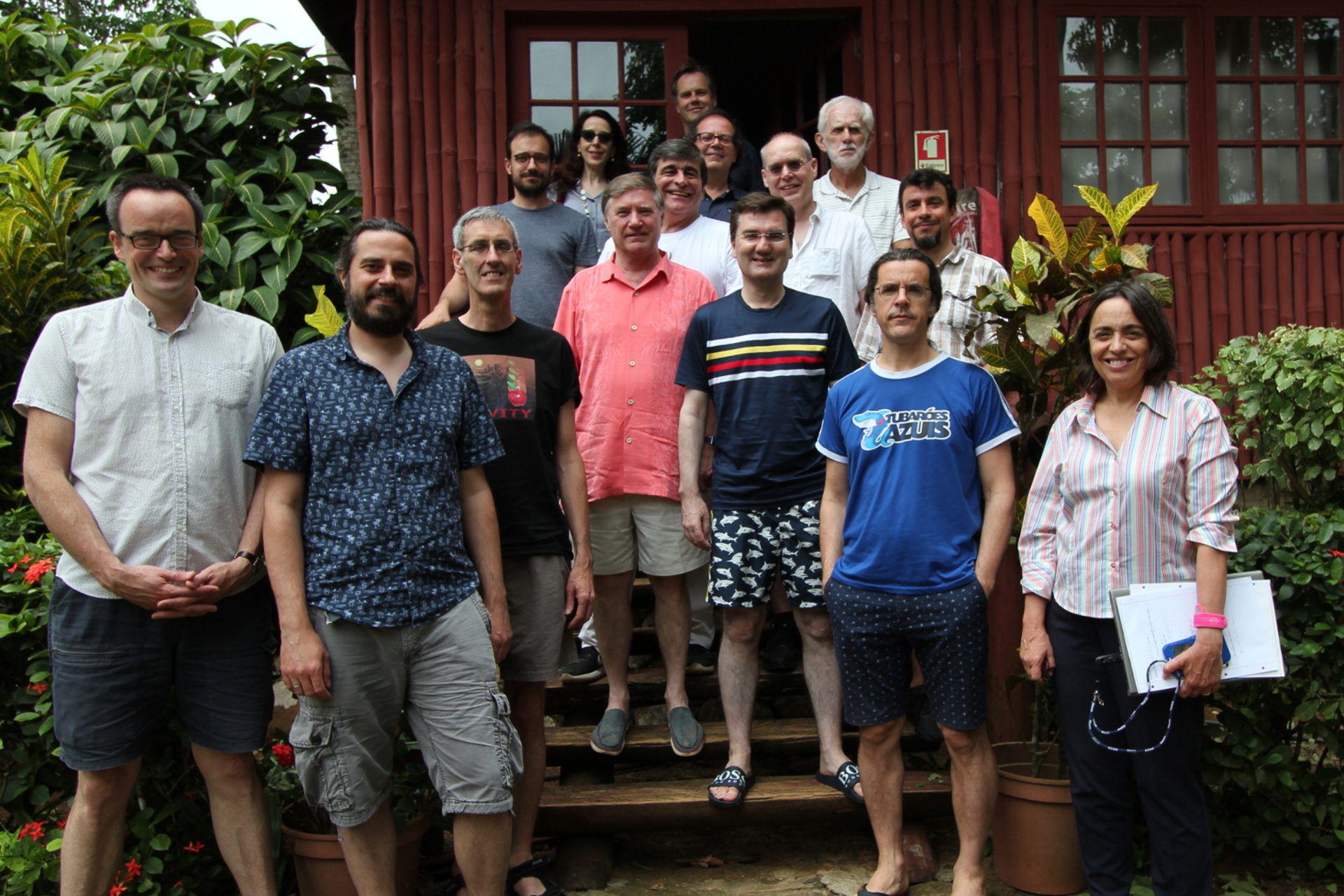}}
\caption{\footnotesize The speakers
of the scientific conference ``From
Einstein and Eddington to LIGO: 100 years of gravitational light
deflection'' in front of the seminar room in resort Bom Bom, Principe.
From top to bottom and from left to right: Pedro Ferreira;
Alessandra Buonanno, Ismael Tereno, Cliff Will; João Costa, José Sande
Lemos, Uli Sperhake; John Barrow, Carlos Herdeiro; Frank Eisenhauer,
Ilídio Lopes; Jonathan Gair, Thomas Sotiriou, Vítor Cardoso, Ana
Mourão. Photograph taken by Jorge Vicente during the
morning coffee break on May 27, 2019.}
\label{speakers}
\end{figure}

On the 27h there were four morning talks where
the themes were experimental and observational tests
of general relativity, gravitational lenses, compact
objects, and numerical relativity. During the coffee break
one could walk through the luxuriant nature and photographs
were taken, see Fig.~\ref{organizers} and
Fig.~\ref{speakers}.
In the afternoon there were four talks about black holes, their
exterior, their interior, and on fundamental properties
of the event horizon. A free discussion ensued
which finished at 7pm. In the evening, Tim de Zeuw of the
Max Planck Institute at Garching, that was also present
in the resort Bom Bom talks, gave a talk for the general public
about the future of astronomy in a reception at Roça Belo Monte.

On the 28th there were seven
talks dedicated to tests of general relativity and
cosmology. In the coffee breaks and in the afternoon debate
there were discussions about the past and future
of astrophysics and gravitation, where the historical
foundation was always present with emphasis on the
creative work of Einstein and Eddington. 
There were conversations about gravitational waves
and what LIGO can still give us and what it is intended
in the future with LISA
(Laser Interferometer Space Antenna)
an ESA project to put satellites in space to detect 
gravitational waves coming from supermassive black
holes and from the primordial universe. 
There was also a debate about unification theories,
that were initiated and promoted by Einstein and
Eddington, and its union with quantum mechanics, 
and also how black holes can elucidate in a correct formulation
of quantum gravitation, a theory yet to be elaborated.
José Sande Lemos and Jonathan Gair recalled Donald Lynden-Bell
from Cambridge University, their supervisor in the years 1980s
and 2000s, respectively, a great admirer of Eddington. He occupied in
the Institute of Astronomy, Eddington's room  with a
famous curved door and where a photograph
of the great
astrophysicist hanged on the wall over the working table.
The works finished at 7pm and there followed
a reception in Casa Rosa, the official house of the governor in
Santo António, where scientists and political representatives 
of São Tomé e
Principe and Portugal participated.

\begin{figure}[h]
{\includegraphics[scale=0.20]{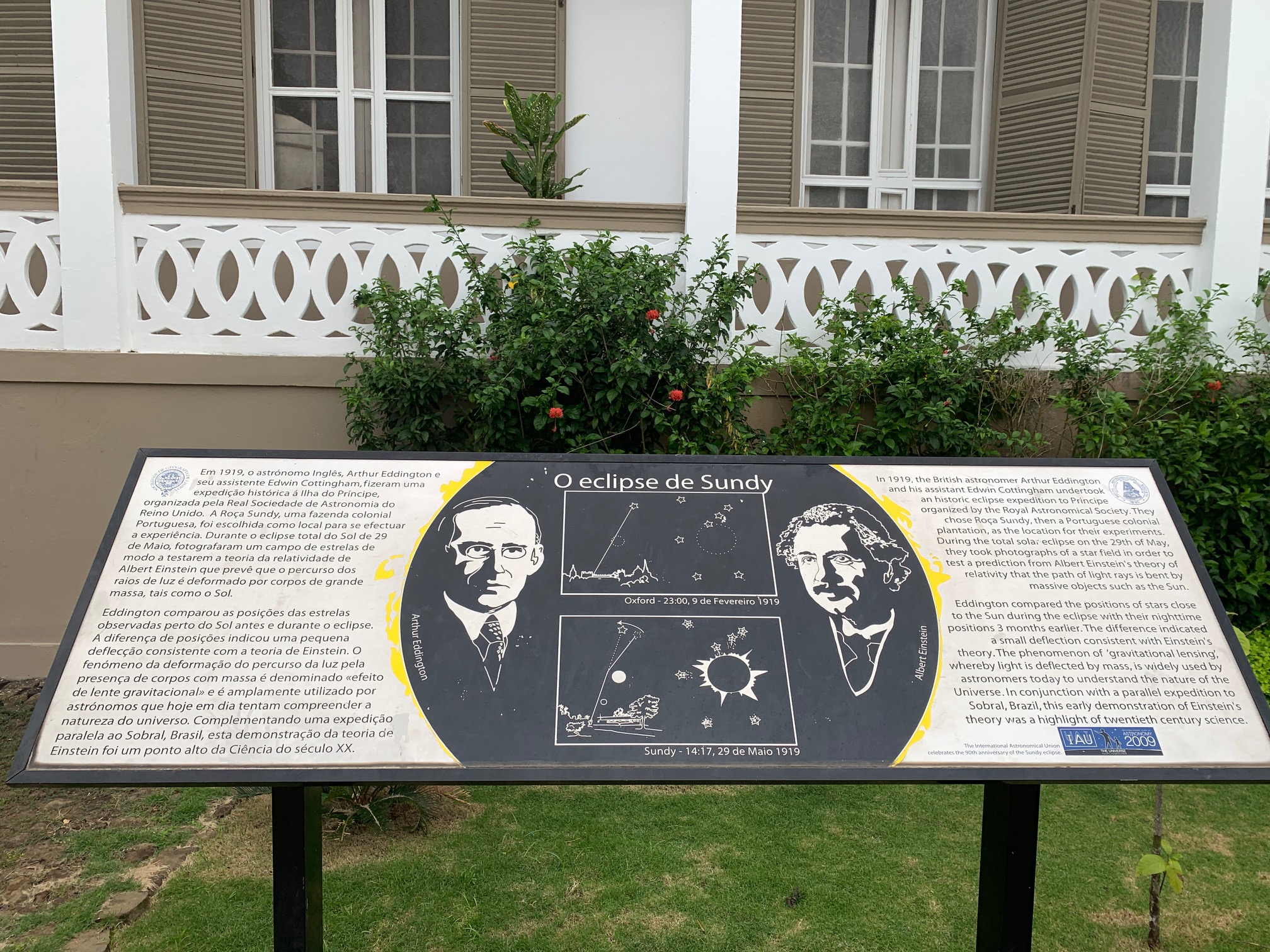}}
\caption{\footnotesize Informative plate in Roça Sundy.
Photograph taken by Ilidio Lopes in May 29, 2019.}
\label{informativorocasundy}
\end{figure}

On the 29th, the participants of the scientific conference
were in Roça Sundy, see Fig.~\ref{informativorocasundy}.
At Sundy there was a public event with special celebrations
exactly 100 years after the eclipse. Of particular relevance, 
the Principe and Sobral celebrations got together in a teleconference
at 2:30pm Principe hour, 10:30am Sobral hour,
for a joint celebration.
The speakers, by this order, were
the Prefect of Sobral Ivo Gomes,
the Prime Minister of São Tomé e Principe Jorge Bom Jesus,
the President of the Regional government of Principe
José Cassandra, the Governor of Ceará Camilo Santana,
the President of the Brazilian Academy of Sciences Luiz
Davydovich, the President of the Brazilian Society for the
Progress of Science Ildeu Moreira, the Rector of the
University of 
São Tomé e Principe Aires Bruzaca de Menezes,
the President of the International
Astronomical Unions Ewine Dishoeck,
and the President of the Center of
Astrophysics and Gravitation of Lisbon and President of
the General Assembly of the Portuguese Society of Relativity
and Gravitation José Sande Lemos. There were congratulations
from all for this special moment. 

The scientific conference in Principe appeared in CENTRA News \cite{centranews},
in IST News \cite{istnews}, and was covered by
the New York Times~\cite{nyt}.
For the history and science of the 1919 eclipse 
see  \cite{jpsl2019}.
There were many celebrations all over the world,  we refer
to some in the following.

\section{Other celebrations in 2019}

\subsection{Eddington at Sundy in Principe}

For the Principe celebrations there was an extensive educational and
scientific project ``Eddington na Sundy: 100 years after'' organized
by the coordinator Joana Latas in cooperation with several entities,
in particular with the Principe Regional Government.  It was a project
with several fronts that is intended to have continuity, see [7].  An
aim of the project was to attract the attention of the Principe
inhabitants to the relevance of the 1919 observations and to science
in general. Local celebrations occurred from May 25 to May 30, 21019,
the high point happening at Roça Sundy on May 29, where during the day
national and international figures were present.  In that day the
population of Principe showed its great hospitality to the hundreds of
participants that came from abroad. An exhibition was opened in Roça
Sundy itself on the detection of light deflection. The exhibition is
now permanent.  Principe and Sobral joined celebrations in a
videoconference. The scientific conference gladly
joined this comprehensive
educational and scientific
project.

\subsection{Sobral}
In Sobral there was a scientific conference and a major public event
from May 26 to May 31, that was in tune with the expectations and 
the importance of the discovery.

\subsection{Lisbon}
A special number of Gazeta de Física, a Portuguese journal that
disseminates and promotes physics in general, was published in May
2019 to celebrate the events in Principe and in Sobral
\cite{jpslfitascrawford,jpslemosherdeirocardosogazeta}.  An exhibition
opened in May 2019 in the National Museum of Natural History and
Science of the University of Lisbon with the title ``E3 - Einstein,
Eddington and the Eclipse''.  The XXIX Astronomy and Astrophysics
National Meeting, this year
organized at Instituto Superior Tecnico, University
of Lisbon, was dedicated to Eddington and the eclipse,
see~\cite{enaaweb}.

\subsection{Rio de Janeiro}

At Rio de Janeiro National Observatory, house of its illustrious
director Henrique Morize, that was present in Sobral to observe the
solar corona and helped the English expedition in many ways, there was
a celebration meeting in May 2019, just before the Sobral event.

\subsection{London}

In London there was a public event on November 6, 2019,
organized by the Royal Astronomical Society
celebrating de historical meeting of November 6, 2019,
that was
presided by J.~J.~Thomson, the man of the electron and president
of the Royal Society at the time, that gathered the two
societies to officially announce the results of the
measurements of the light deflection by 
 Dyson, Crommelin, Davidson, and Eddington, that
 confirmed Einstein's theory of gravitation.

\subsection{Future}
We hope that this May 29
date be always commemorated, with particular
emphasis at each 100 years from 1919 onward, as we have done now
for the first time, and that Einstein, Eddington, Principe, and
Sobral be remembered in this date.
It will  show that gravitation theory, realized in general relativity
or eventually in some other more fundamental theory,
continues prosperous.


\vskip 1.0cm
\centerline{\bf Acknowledgments}

\vskip 0.3cm
We thank Phillipe Moreau and Beatriz Geraldes from Hbd Stp -
Investimentos Turisticos 
for their kind assistance in the handling
of the scientific conference in resort Bom Bom,
and Nuno Santos and José Quina, managers
at the resort, for all the help during the
conference. We thank Joana Latas, coordinator
of the Eddington at Sundy organization, for all
the help before and during our scientific conference.
We thank Dulce Conceição of CENTRA for dealing
with all the administrative processes for the
conference and 
Sérgio Almeida for the elaboration
of the conference webpage. We thank CENTRA and its members
for the complete support for the realization of the
scientific conference.
We thank Instituto Superior T\'ecnico, and in particular
our colleague Luis Viseu Melo, for all the
help and simplifications
in the financial and administrative processes. 

We thank Fundação para a Ciência e Tecnologia (FCT), Portugal,
for the financial help through the
project~No.~UID/FIS/00099/2019.

\end{document}